\documentclass[twocolumn,showpacs,preprintnumbers,amsmath,amssymb,superscriptaddress]{revtex4}

\usepackage{graphicx}% Include figure files
\usepackage{dcolumn}% Align table columns on decimal point
\usepackage{bm}% bold math
\usepackage{longtable}% bold math

\begin{document}

\preprint{APS/123-QED}

\title{Random Walks on deterministic Scale-Free networks: Exact results}

\author{E. Agliari}
\affiliation{Dipartimento di Fisica, Universit\`a degli Studi di
Parma, viale Usberti 7/A, 43100 Parma, Italy}
\author{R. Burioni}
\affiliation{Dipartimento di Fisica, Universit\`a degli Studi di
Parma, viale Usberti 7/A, 43100 Parma, Italy}
\affiliation{INFN, Gruppo Collegato di
Parma, viale Usberti 7/A, 43100 Parma, Italy}

\date{\today}

\begin{abstract}
We study the random walk problem on a class of deterministic
Scale-Free networks displaying a degree sequence for hubs scaling as a power
law with  an exponent $\gamma=\log 3/\log2$.
We find exact results concerning different first-passage phenomena
and, in particular, we calculate the probability of first return
to the main hub. These results allow to derive the exact
analytic expression for the mean time to first reach the main hub, whose
leading behavior is given by $\tau \sim V^{1-1/\gamma}$, where
$V$ denotes the size of the structure, and the mean is over a set
of starting points distributed uniformly over all the other sites
of the graph. Interestingly, the process turns out to be
particularly efficient. We also discuss the thermodynamic limit of
the structure and some local topological properties.
\end{abstract}

 \maketitle

\section{\label{sec:intro}Introduction}

In the last few years, one of the most studied topic in network theory has
been the investigation of highly inhomogeneous structures. Networks with
strong variability in local metric and topological properties have been
shown to well represent many real structures, occurring in nature and in
man-made systems. Condensed matter and soft materials often
feature an inhomogeneous organization in space \cite{comb},
and even engineered devices can be constructed to reproduce a highly
varying arrangement, in order to obtain the designed physical properties
\cite{Levy}. Such networks, being physically embedded in space, are finite
dimensional and it is known that in this case inhomogeneous topology can strongly
affects physical phenomena occurring on the network itself.

Graphs are also used to describe the generic relation between a set of
elements or agents, as it happens in complex networks theory in biology,
social science, computer science and economy \cite{AB,dor}. Then, these
networks often feature infinite dimensionality, and inhomogeneity has been
detected in several real systems \cite{blumen,blumen2,gallos,condamin}.
One of the most studied structures in the literature are Scale-Free (SF)
networks, which show a degree sequence scaling with a characteristic power
law \cite{AB,fal}. This implies that highly inhomogeneous regions can be
present in the network.

Random structures  and stochastic approaches have appeared to be very useful in this framework, even if, in real cases, one always has to deal with a single realization of the disorder. Due to the strong variability of the topology in the samples, the quenched properties  could not coincide with the  behavior found in the typical cases, as described by the probability distribution.
Therefore, the study of quenched samples and deterministic topologies is a very interesting task \cite{zhang,zhang2,haynes,kozak2,bollt,kittas}.

Apart from the discussion on how and why structures with scaling degree sequences are so often encountered in nature \cite{doyle1}, they certainly represent a new and interesting class  of  inhomogeneous graphs. Original techniques have been developed in order to characterize these topologies and their effect on physical properties. However, the general topological and metric features of SF graphs, on the global and on the local scale, are still not completely understood.

Random walks are one of the simplest stochastic processes affected by the topology of a network and, at the same time, the basic model of diffusion phenomena and non-deterministic motion.
They have been extensively studied for decades on regular lattices \cite{weiss}, fractal networks and finite dimensional inhomogeneous structures \cite{havlin}, where they have been shown to be able to evidence
a new and unexpected phenomenon arising in presence of strong inhomogeneity, namely the splitting between local and average properties \cite{burioni}.
The richer topology of a generic, inhomogeneous and infinite dimensional graph can have a dramatic effect on the properties of random walks, especially when considering infinite graphs, which are introduced to describe macroscopic systems in the thermodynamic limit.
Random walks represent  not only a good model for  diffusion phenomena on large complex networks, but also a direct way to characterize their large scale topological features, also in presence of strong inhomogeneity, and their influence on physical properties. Once these are known, they could also be used to fruitfully design and engineer a network  topology with given properties.

In this paper we want to deepen the analysis by studying random walks on a specific Scale-Free topology, namely a deterministic Scale-Free network, built in a recursive way and featuring a scaling distribution for the hub degrees, with an exponent $\gamma= \log3/\log2$. For this deterministic structure, first introduced in \cite{barabasi}, some metrical and spectral properties have recently been investigated in details \cite{iguchi}. On the other hand, how these properties are linked to diffusion processes on the network is still an open problem.

Using the formalism of random walks generating functions
\cite{weiss}, we derive exact expressions for the first passage
times \cite{redner} for a random walker starting from the
maximally connected hub and from the last generation ``rims" of
the deterministic SF network. In particular, we investigate their
dependence on the size of the network and  we derive an exact
expression for the mean time to first reach the most connected
hub, namely the mean time to absorption if we place a perfect trap
on the most connected hub. This quantity displays an extremely
slow behavior as the size of the network increases, given by
$\tau \sim V^{1-1/\gamma}$, where $V$ denotes the size of the
structure, and the mean is over a set of starting points
distributed uniformly over all the other sites of the graph.
Interestingly, the trapping process appears to be very efficient.
We also obtain an implicit relation for the generating function of
the first passage probabilities from the hub to rims, and
viceversa, which provides some insight into the leading
singularity for the generating function of the return probability
to the hub. In particular, we prove the recurrence of
the main hub by exploiting the connection between random walks and
electric networks \cite{doyle}.
Moreover, we compare the return probability to the main hub with the return probability to an end-node evidencing a very different asymptotic behavior.
All the results are checked with extensive numerical calculations.

The paper is organized as follows:
In Sec.~\ref{sec:deterministic} we give a brief mathematical description of the deterministic Scale-Free graph, introducing the language and the formalism we will use in the whole article. Then, in Sec.~\ref{sec:Res} we derive the recursive relation between the first passage probabilities from hubs to rims and vice versa; from these we calculate the exact expression for the mean time to first reach the most connected hub and its dependence on the volume of the network. In Sec.~\ref{sec:Conclusions}  we calculate the return probability on the main hub and we discuss its recurrence in the thermodynamic limit. Finally, Sec.~\ref{sec:Discussion} is devoted to conclusion and discussion.

\section{\label{sec:deterministic}Deterministic Scale-Free network}
\begin{figure}[tb] \begin{center}
\includegraphics[width=.55\textwidth]{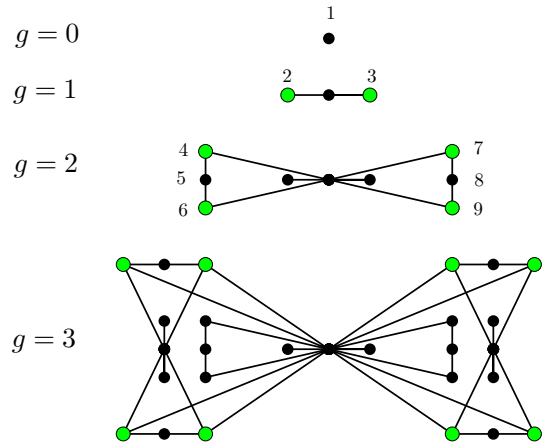}
\caption{\label{fig:grafo} (Color on line) Iterative construction of the deterministic scale-free network; the first four generations are depicted. Nodes belonging to the set $\mathcal{B}_g$ are represented in brighter color.}
\end{center}
\end{figure}

A generic graph $\mathcal{G}$ consists in a non-empty set $\mathcal{V}$ of nodes joined pairwise by a set of links $\mathcal{L}$ \cite{harary}. Here we consider a particular set of deterministic graphs $\{ \mathcal{G}_g \}_{g=0,1,2,...}$, first introduced in \cite{barabasi}, which can be built recursively: At the $g$-th iteration one has the graph of generation $g$, denoted by  $\mathcal{G}_g$ (see Fig.~\ref{fig:grafo}).

Starting from the so called root constituted by one single node labeled as $i=1$, at the first iteration one introduces two more nodes $i=2, 3$ and connects each of them to the root; the resulting chain of length three represents the graph of generation $g=1$. We call $\mathcal{B}_1=\{2,3\}$ the set of sites added at the first generation and linked to the root. Then, at the second iteration one adds two chains of length three and connects each end node to the root, namely $\mathcal{B}_2=\{4,6,7,9\}$, so that the root will increase its coordination number from $2$ to $6$. Proceeding analogously, at the $g$-th iteration one introduces two replica of the existing graph of generation $g-1$ and connects the root with each site making up $\mathcal{B}_g$.

Hence, at the $g$-th generation the root turns out to be the main
hub with coordination number $2(2^{g}-1)$, the set of all nodes
has cardinality $V_g \equiv |\mathcal{V}_g|=3^g$ and
$|\mathcal{B}_g|=2^g$. As shown in \cite{barabasi},
$\mathcal{G}_g$ exhibits $(2/3)3^{g-i}$ ``hubs'' with degree
$2^{i+1}-2$, being $i \in [1,g-1]$. As a result, the tail of the
degree distribution is a power law $P(k) \sim k^{-\gamma}$ with
exponent $\gamma=\log 3 / \log2 \approx 1.59$. However, this does
not hold for the so called ``rims'' contained in the sets
$\mathcal{B}_g$. In this case one finds $P(k) \sim (2/3)^k =
e^{-\bar{\gamma} k}$, where $\bar{\gamma} = \log(3/2)\approx 0.405
$, which shows that the scaling nature of the rims is not
scale-free but exponential \cite{iguchi}. The topological and
spectral properties of this graph have been deeply analyzed in
\cite{iguchi} where, in particular, the average degree is shown to
be $\langle k \rangle_g \equiv \sum k P(k) / V_g = 4 [1-(2/3)^g]$,
namely it approaches $4$ as $g \rightarrow \infty$. Indeed, on the
one hand the graph becomes very complex and there exists a set of
few [$o(1)$] nodes whose coordination number grows indefinitely,
on the other the number of nodes with degree $\leq 2$ become large
indefinitely; as a result the average degree remains finite, conversely the second moment $\langle k^2 \rangle$ is divergent.
It should also be underlined that by increasing $g$ the number of cycles grows fast; as a result, while the number of end-nodes grows linearly with the volume $V$ ($\sim 3^g$), the average distance from the main hub increases only logarithmically with $V$ ($\sim g$).

\section{\label{sec:Res} Generating functions and first passage properties}

The simple random walk (RW) on a graph $\mathcal{G}$ is defined by the jumping probability $p_{ij}$ between nearest neighbor sites $i$ and $j$:
\begin{equation} \nonumber
p_{ij}=\frac{A_{ij}}{z_i} = (\mathbf{Z}^{-1}\mathbf{A})_{ij},
\end{equation}
where $Z_{ij}=z_i \delta_{ij}$ and $z_i=\sum_{i \in \mathcal{V}}A_{ij}$ is the coordination number of site $i$. Therefore, the probability of reaching in $t$ steps site $j$ starting from $i$ is
\begin{equation} \label{eq:power_A}
P(i,j;t)=(p^t)_{ij}.
\end{equation}

In the following we denote with $P_g(i,j;t)$ the probability that a RW on $\mathcal{G}_g$, starting from a site $i$ reaches the site with label $j$ at time $t$ and $F_g(i,j;t)$ the probability that the same walk reaches the site $j$ for the first time at time $t$. Moreover, we consider the probability $B_g(t)$ that, at the generation $g$, a walk starting from any site in $\mathcal{B}_g$ first reaches the central hub and the probability $H_g(t)$ that a walk starting from the central hub first reaches any site in $\mathcal{B}_g$.

Now, the following equations hold
\begin{equation}\label{eq:B}
B_g(t)=\frac{\delta_{t,1}}{g} + \frac{1}{g} \sum_{l=1}^{g-1} \sum_{k=1}^{t-1} H_l(k) B_g(t-1-k)
\end{equation}
and
\begin{equation}\label{eq:H}
H_g(t)=\frac{2^{g-1}}{2^g-1} \delta_{t,1} + \sum_{l=1}^{g-1} \sum_{k=1}^{t-1} \frac{2^{l-1}}{2^g-1}  B_l(k) H_g(t-1-k).
\end{equation}
Some comments are in order here. In a graph of generation $g$,
each rim is connected to $g$ nodes of which one is the main hub;
this accounts for the first term in Eq.~\ref{eq:B}. The remaining
$g-1$ links connect each rim to a minor hub from which one can
reach the main hub only passing through a rim; this explains the
sum of convolutions in Eq.~\ref{eq:B}. Analogously, from the main
hub $2^g$ links out of $2(2^g-1)$ point directly to a rim; the
remaining links connect the main hub to nodes corresponding to
rims of graphs of generations $l<g$ and from such nodes one can
reach any site in $\mathcal{B}_g$ only through the main hub
itself.

The generating functions corresponding to Eq.~\ref{eq:B} and Eq.~\ref{eq:H} read as
\begin{equation}\label{eq:B_gen}
\tilde{B}_g(z) \equiv \sum_{t=0}^{\infty} B_g(t)z^t=\frac{z}{g}+\frac{z}{g}\sum_{l=1}^{g-1}\tilde{H}_l(z) \tilde{B}_g(z)
\end{equation}
and
\begin{equation}\label{eq:H_gen}
\tilde{H}_g(z) \equiv \sum_{t=0}^{\infty} H_g(t)z^t =\frac{2^{g-1}}{2^g-1}z+ \frac{\tilde{H}_g(z)}{2^g-1}z \sum_{l=1}^{g-1} 2^{l-1}\tilde{B}_l(z).
\end{equation}
Notice that, as can be easily inferred from Fig.~\ref{fig:grafo}, $B_1(t)=H_1(t)=\delta_{t,1}$, from which $\tilde{B}_1(z)=\tilde{H}_1(z)=z$. We also recall that, by definition, $\tilde{B}_g(1)$ is just the probability to ever reach the hub from any site in $\mathcal{B}_g$ and, analogously, $\tilde{H}_g(1)$ is the probability to ever reach any site in $\mathcal{B}_g$ from $i=1$. Of course, for finite $g$ one has $\tilde{B}_g(1)=\tilde{H}_g(1)=1$.
With some algebra we rewrite Eq.~\ref{eq:B_gen} and Eq.~\ref{eq:H_gen} respectively as
\begin{eqnarray}\label{eq:B_gen_2}
\tilde{B}_g(z) \left( \frac{g}{z} -  \sum_{l=1}^{g-1}\tilde{H}_l(z) \right) =  1\\
\label{eq:H_gen_2} \tilde{H}_g(z) \left( \frac{2^{g}-1}{2^{g-1} \;
z} - \sum_{l=1}^{g-1} 2^{l-g} \tilde{B}_l(z) \right) = 1.
\end{eqnarray}
Moreover, by properly handling Eq.~\ref{eq:B_gen_2} and Eq.~\ref{eq:H_gen_2} we can obtain the following two finite-difference equations coupled together:
\begin{eqnarray}\label{eq:sistema1}
\frac{1}{\tilde{B}_{g+1}(z)}-\frac{1}{\tilde{B}_g(z)}=\frac{1}{z} - \tilde{H}_g(z)\\
\label{eq:sistema2}
\tilde{B}_g(z) - \frac{2}{z} = - \frac{2}{\tilde{H}_{g+1}(z)} + \frac{1}{\tilde{H}_g(z)},
\end{eqnarray}
which can be combined together to get the rather symmetric expression $\tilde{B}_g(z) \left\{ [\tilde{B}_{g+1}(z)]^{-1} - [\tilde{B}_{1}(z)]^{-1} \right\} = 2 \tilde{H}_g(z) \left\{ [\tilde{H}_{g+1}(z)]^{-1} - [\tilde{H}_{1}(z)]^{-1} \right\}$.

It is interesting to notice that, since $0 \leq \tilde{H}_g(z)
\leq 1$, for any $g$ and for any $z$, from Eq.~\ref{eq:B_gen_2} we have $z/g
\leq \tilde{B}_g(z) \leq z/[g(1-z)+z]$, from which one gets
$\tilde{B}_{\infty}(z)=0$. Hence, in the thermodynamic limit the
probability to eventually reach the hub from $\mathcal{B}_g$ is zero.

From Eq.~\ref{eq:B_gen_2} and Eq.~\ref{eq:H_gen_2} it is possible to calculate recursively $\tilde{B}_g(z)$ and $\tilde{H}_g(z)$; for instance, for the first generations we get
\begin{eqnarray}\label{eq:esempi_B}
\tilde{B}_2(z)&=&\frac{z}{2-z^2} ,\\
\tilde{B}_3(z)&=&\frac{z(3-z^2)}{9-8z^2+z^4},\\
\tilde{B}_4(z)&=&\frac{z(3-z^2)(14-11z^2+z^4)}{168-282z^2+145z^4-24z^6+z^8},
\end{eqnarray}
and
\begin{eqnarray}\label{eq:esempi_H1}
\tilde{H}_2(z)&=& \frac{2z}{3-z^2}, \\
\label{eq:esempi_H2}
\tilde{H}_3(z)&=& \frac{4z(2-z^2)}{14-11z^2+z^4}, \\
\label{eq:esempi_H3}
\tilde{H}_4(z)&=& \frac{8z(2-z^2)(9-8z^2+z^4)}{270-435z^2+211z^4-31z^6+z^8}.
\end{eqnarray}

The generating functions $\tilde{B}_g(z)$ and $\tilde{H}_g(z)$ allow to calculate the average time $t_g^B$ taken by a random walk started on a site in $\mathcal{B}_g$ to reach the hub and the average time $t_g^H$ taken by a random walk started on the main hub to reach any site in $\mathcal{B}_g$, respectively. In fact, for any arbitrary generation $g$ we can write:
\begin{eqnarray}
t_g^B = \frac{\partial}{\partial z} \tilde{B}_g(z)\bigg|_{z=1}\\
t_g^H = \frac{\partial}{\partial z} \tilde{H}_g(z)\bigg|_{z=1}.
\end{eqnarray}
Hence, deriving Eq.~\ref{eq:B_gen_2} and Eq.~\ref{eq:H_gen_2} and recalling that for finite structures $\tilde{B}_g(1)=\tilde{H}_g(1)=1$, we find the following coupled equations:
\begin{eqnarray}\label{eq:uno}
t_g^B &=& g + \sum_{l=1}^{g-1}t_l^H, \\
\label{eq:due}
t_g^H &=& 2 - 2^{1-g} + 2^{-g}\sum_{l=1}^{g-1}2^l t_l^B.
\end{eqnarray}
Notice that the previous two equations could be obtained directly from Eq.~\ref{eq:B} and Eq.~\ref{eq:H} by multiplying both sides by $t$ and summing over $t=0,...,\infty$.

Now, from Eq.~\ref{eq:uno} and Eq.~\ref{eq:due} we get
\begin{eqnarray}
t_{g+1}^B - t_g^{B} &=& 1 + t_g^{H} \\
2 t^H_{g+1} -  t_g^{H} &=& 2 + t_g^B.
\end{eqnarray}
Such equations can be properly handled to get
\begin{equation}\label{eq:rec_H}
2t^H_{g+2} - 3t_H^{g+1}=1,
\end{equation}
and
\begin{equation}\label{eq:rec_B}
2t^B_{g+2} - 3t^B_{g+1} =3,
\end{equation}
whose solution is given by
\begin{equation}\label{eq:t_H}
t_g^H=\frac{4}{3}\left( \frac{3}{2}\right)^g -1,
\end{equation}
and
\begin{equation}\label{eq:t_B}
t_g^B=\frac{8}{3}\left( \frac{3}{2}\right)^g -3,
\end{equation}
where we used $t_1^H=t_1^B=1$.

Interestingly, $t_g^B$ and $t_g^H$ both display the same exponential law: The average time to first reach the hub from any of the $2^g$ sites making up $\mathcal{B}_g$ and the average time to first reach any site in $\mathcal{B}_g$ from the hub grows with the generation as $\sim (3/2)^g $. Also, recalling that $|\mathcal{B}_g|=2^g$ and that $V_g = 3^g$, we can write $2^g=V^{\log 2/ \log 3}$ and get $(3/2)^g = V_g/|\mathcal{B}_g| = V^{1-\log2/\log3}$.

The average times $t_g^B$ and $t_g^H$ found above are useful to calculate the mean time to absorption $\tau_g$. In fact, let us define $t_g(i)$ the mean time necessary to first reach the hub from site $i \neq 1$, then $\tau_g \equiv \sum_{i \in \mathcal{V'}_g} t_g(i) / (V_g -1)$ and we can write
\begin{equation}
\sum_{i \in \mathcal{V'}_g} t_g(i) = \sum_{i \in \mathcal{V'}_{g-1}} t_{g-1}(i) + \sum_{i \in \mathcal{V}_g \setminus \mathcal{V}_{g-1}} t_g(i),
\end{equation}
namely
\begin{equation}\label{eq:somme}
\tau_g = \tau_{g-1} \frac{V_{g-1}-1}{V_g-1} + \frac{1}{V_g-1} \sum_{i \in \mathcal{V}_g \setminus \mathcal{V}_{g-1}} t_g(i),
\end{equation}
where $\mathcal{V'}_g \equiv \mathcal{V}_g \setminus \{ 1 \}$ and $\mathcal{V}_g \setminus \mathcal{V}_{g-1}$ is the set of sites added at the $g$-th generation.
We now notice that the last sum in Eq.~\ref{eq:somme} is just given by
\begin{eqnarray}
\nonumber
\sum_{i \in \mathcal{V}_g \setminus \mathcal{V}_{g-1}} t_g(i) = t_g^B |\mathcal{B}_g| + (t_g^B +1) \frac{|\mathcal{B}_g|}{2} \\
+ \sum_{l=1}^{g-2}\left[ \tau_g(3^l-1)+3^l t_{l+1}^H+3^lt_g^B \right ]2^{g-l-1},
\end{eqnarray}
where the first two terms simply allow for $|\mathcal{B}_g|$ rims and $|\mathcal{B}_g|/2$ nodes bridging between two rims; then, from the remaining nodes making up $\mathcal{V}_g \setminus \mathcal{V}_{g-1}$ one can reach the main hub by passing through any rim, possibly via a minor hub, and this yields to the last sum.

By replacing in the previous equation the expressions for $t_g^B$
and $t_g^H$ found in Eqs.~\ref{eq:t_H}, \ref{eq:t_B} we get
\begin{equation}
\sum_{i \in \mathcal{V}_g \setminus \mathcal{V}_{g-1}}
t_g(i)=\frac{2^g}{5}+\frac{32}{15} \left (\frac{9}{2} \right) ^{g}
- \frac{8}{3} 3^{g} + 2^{g-1} \sum_{l=1}^{g-2}
\tau_{l}\frac{3^l-1}{2^{l}}
\end{equation}
which, together with Eq.~\ref{eq:somme} yields
\begin{equation}\nonumber
(3^{g+1}-1)\tau^{g+1}-3(3^g-1) \tau^{g} = \frac{16}{3} \times
3^{g} \left[ \left( \frac{3}{2} \right)^g - \frac{1}{2} \right].
\end{equation}
The solution of this recursive equation is
\begin{equation} \label{eq:final}
\tau_g = \frac{1}{3^g-1}\left[ \frac{32}{9} \left( \frac{9}{2} \right)^g - \frac{2}{9} (17 + 4g)  3^g  \right],
\end{equation}
whose leading behavior is given by
\begin{equation}  \label{eq:final_leading}
\tau^g \sim V_g^{1- \log 2 / \log3} = V_g^{1-1/\gamma},
\end{equation}
with $1- \log 2 / \log3 \approx 0.37$. Such an exponent is even lower than the exponent $\log 2 / \log 3 = 1/(\gamma'-1)\approx 0.63$ found in \cite{zhang} for a deterministic scale-free network displaying the power degree $\gamma' =1+ \log 3/ \log2 = 1 + \gamma$. Indeed, the multi-fractal nature displayed by the graph under study gives rise to a remarkable efficiency for trapping on $i=1$.

\begin{figure}[tb] \begin{center}
\includegraphics[width=.5\textwidth]{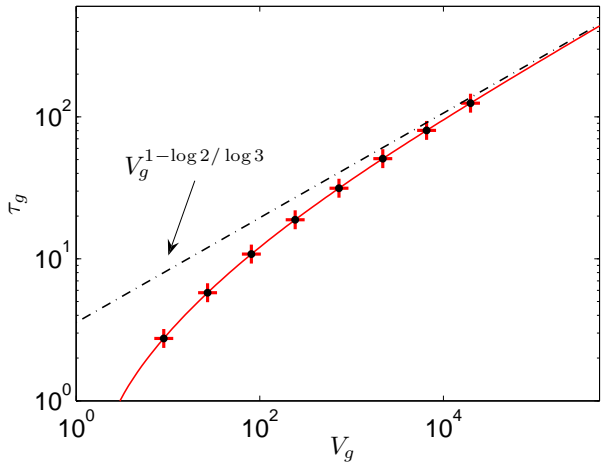}
\caption{\label{fig:compare} (Color on line) Mean first-passage
time $\tau^g$ for a simple random walker moving on the
deterministic sale-free network as a function of the volume $V_g$;
data from the exact analytic expression in Eq.~\ref{eq:final}
($\bullet$) are compared to the asymptotic form of
Eq.~\ref{eq:final_leading} (dash-dotted line) and to the numerical
solution of Eq.~\ref{eq:tau_av} ($+$, the continuous line is a
guide to the eye).}
\end{center}
\end{figure}

\subsection{\label{sec:Numerical}Numerical calculations}
For a generic graph, given the corresponding adjacency matrix $\mathbf{A}$ and the coordination matrix $\mathbf{Z}$, the numerical calculation of the mean time to absorption can be performed by exploiting a differential equation where the normalized discrete Laplacian $\mathbf{\Delta} = \mathbf{A} \mathbf{Z}^{-1} - \mathrm{\mathbf{I}}$ appears \cite{kozak,agliari,zhang,zhang2}. More precisely, for the topological structures analyzed here, the Laplacian $\mathbf{\Delta}_g$ is a $V_g \times V_g$ matrix which depends on the generation $g$ and we have
\begin{equation}
- \sum_{j=2}^{V_g} \mathbf{\Delta_g}_{ij} t_g(j) = 1.
\end{equation}
Therefore, the mean time to absorption averaged over all possible starting sites $i \neq 1$ reads as:
\begin{equation}\label{eq:tau_av}
\tau_{g} = \frac{1}{V_g-1} \sum_{i=2}^{V_g} t_g(i) =
\frac{1}{V_g-1} \sum_{i=2}^{V_g}\sum_{j=2}^{V_g}
(-\mathbf{\Delta_g}^{-1})_{ij}.
\end{equation}
In Fig.~\ref{fig:compare} we compare the analytical results of Eq.~\ref{eq:final} and of Eq.~\ref{eq:final_leading} with the numerical results obtained via Eq.~\ref{eq:tau_av}.

\section{\label{sec:Conclusions} Return probability}
The results found in the previous section allow to deepen the analysis of the random walk problem on the determinist structure considered. In particular, the probability $F_g(1,1;t)$ that a random walk started on the hub returns to the hub itself for the first time after time $t$ has the form
\begin{equation}\label{eq:first_return}
F_g(1,1;t)=\sum_{i=1}^{g} \frac{|\mathcal{B}_i|}{\mathbf{Z_g}_{11}} B_i(t-1) = \sum_{i=1}^{g} \frac{2^i}{2(2^g-1)} B_i(t-1)
\end{equation}
and the pertaining generating function is
\begin{equation}\label{eq:first_return_gen}
\tilde{F}_g(z) \equiv \sum_{t=0}^{\infty} F_g(1,1;t) z^t = z
\sum_{i=1}^{g} \frac{2^{i-1}}{2^g-1} \tilde{B}_i(z).
\end{equation}
By replacing the expression for $\tilde{B}_g(z)$ appearing in Eq.~\ref{eq:sistema2}, we get the telescopic sum $\sum_{i=1}^{g} [2^{i+1}/\tilde{H}_{i+1}(z) - 2^{i}/\tilde{H}_{i}(z)]$, so that Eq.~\ref{eq:first_return_gen} simplifies into
\begin{eqnarray}\label{eq:first_return_gen2}
\nonumber
\tilde{F}_g(z) &=& 2 + \frac{z}{2(2^g-1)} \left[ \frac{2}{\tilde{H}_1(z)} -  \frac{2^{g+1}}{\tilde{H}_{g+1}(z)} \right] \\
&=& 2 + \frac{1}{2^g-1} - \frac{2^g z}{(2^g-1) \tilde{H}_{g+1}(z)},
\end{eqnarray}
which highlights that the probability to first return on the hub at generation $g$ directly depends on the probability to first reach any site in $\mathcal{B}_{g+1}$ starting from $i=1$.
We also notice that for finite $g$, $\tilde{F}_g(1)=1$, that is the hub is a recurrent point, as expected for any point on finite graphs \cite{burioni}.

Interestingly, from Eq.~34 we can explicitly derive the average
time $t_g^O$ to first return to the main hub as a function of
$t_g^H$:
\begin{equation}\label{eq:first_return_time}
t_g^O = \frac{\partial}{\partial z} \tilde{F}_g(z) |_{z=1} = \frac{t_{g+1}^H-1}{1-2^{-g}} = \frac{2(3^g - 2^g)}{2^g - 1},
\end{equation}
where in the last equality we used the result of Eq.~\ref{eq:t_H}.
Therefore, for large structures we have $t_g^O \sim (3/2)^g$,
which is the same leading behavior found for $t_g^B$ and $t_g^H$.
This result is also consistent with Kac formula \cite{kac}
according to which $t_g^O = [P(1,1;\infty)]^{-1}= 2
|\mathcal{L}_g|/\mathbf{Z_g}_{11}= \sum_{i \in \mathcal{V}_g}
\mathbf{Z_g}_{ii}/\mathbf{Z_g}_{11}$.

\begin{figure}[tb] \begin{center}
\includegraphics[width=.4\textwidth]{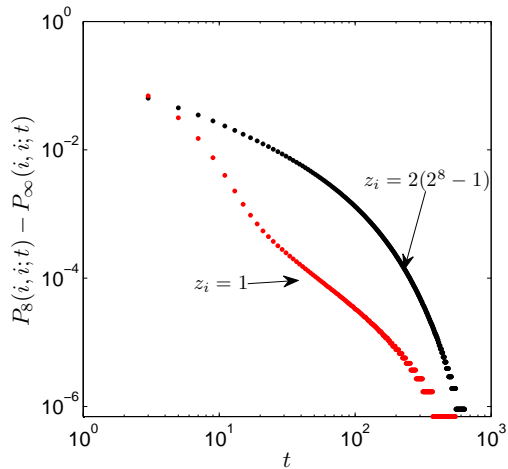}
\caption{\label{fig:p_t} (Color on line) Normalized return probability
$P_g(i,i;t)-P_{\infty}(i,i;t)$ for a random walk on
$\mathcal{G}_8$ and started at the main hub ($i=1$) and at an
end-node belonging to $i \in \mathcal{V}_g \setminus
\mathcal{V}_{g-1}$, as a function of time $t$; only even instants
of time have been depicted.}
\end{center}
\end{figure}

From $\tilde{F}_g(z)$ we can now calculate $\tilde{P}_g(z) \equiv \sum_{t=0}^{\infty} P_g(1,1;t)z^t = [1-\tilde{F}_g(z)]^{-1}$ \cite{burioni}: using Eq.~\ref{eq:B_gen_2} and Eq.~\ref{eq:first_return_gen2} we get
\begin{equation}\label{eq:return_prob}
\tilde{P}_g(z)= \frac{(2^g-1)\tilde{H}_{g+1}(z)}{2^g [z-\tilde{H}_{g+1}(z)]}.
\end{equation}
From Eqs.~\ref{eq:esempi_H1}-\ref{eq:esempi_H3} we can derive
\begin{eqnarray}\label{eq:esempi_P}
\tilde{P}_1(z)&=& \frac{1}{2} \frac{2}{1-z^2}, \\
\tilde{P}_2(z)&=& \frac{3}{4} \frac{4(2-z^2)}{(1-z^2)(6-z^2)} ,\\
\tilde{P}_3(z)&=& \frac{15}{16}
\frac{8(2-z^2)(9-8z^2+z^4)}{(1-z^2)(126-109 z^2+22z^4-z^6)}.
\end{eqnarray}
We checked up to generation $g=8$ that $\tilde{P}_g(z)$ can be written as
\begin{equation}\label{eq:return_prob_general}
\tilde{P}_g(z)=(2^g-1)\frac{f(z)}{(1-z^2)g(z)},
\end{equation}
where $f(z)$ and $g(z)$ are even polynomial of degree $2^g-2$ (in
fact all possible cycles have even length), both devoid of any
factor $(1-z^2)$ and satisfying $f(1)/g(1)=1/(3^g-2^g)$.
In Fig.~\ref{fig:p_t} we show data for $P_8(1,1;t)$ and for
$P_8(i,i;t)$, being $i \in \mathcal{V}_g \setminus
\mathcal{V}_{g-1}$ and $z_i=1$, both obtained numerically from
Eq.~\ref{eq:power_A}: $P_g(i,i;t)=[(\mathbf{\Delta_g} +
\mathbf{I})^{t}]_{ii}$. Interestingly, the decay displayed by the
two probabilities considered is significantly different, which provides a further evidence of the strong inhomogeneity of the graph.

Once the thermodynamic limit is taken, a divergence in
$\tilde{P}_g(z)$ for $z
\rightarrow 1$ would imply that the main hub is recurrent
\cite{burioni}. Recurrence can alternatively be proven
by exploiting the connection between random walks an electric
networks. The escape probability can be determined by
calculating the effective resistivity $R_{eff}$ at fixed distance
from the hub, when all the links of the network are replaced by
unit resistors and a unit voltage is applied between the hub and
the points at fixed distance from it. Then, taking the
thermodynamic limit, the relation between the escape probability
from the hub and the effective resistivity of the network reads
\cite{doyle}:
\begin{equation}
P_{esc}=\lim_{g\to\infty}{1\over \mathbf{Z_g}_{11} R_{eff}(g)},
\end{equation}
where, we recall, $\mathbf{Z_g}_{11}=2(2^g-1)$ is the coordination
number of the hub at generation $g$. Therefore, if
$\mathbf{Z_g}_{11}R_{eff}(g)$ diverges in the thermodynamic limit,
the hub is recurrent, while if it converges to a finite value,
there is a finite probability of escape from the hub. On highly
inhomogeneous graphs such as our scale-free networks, however, a
vanishing $P_{esc}$ may not necessarily imply that the mean number
of visits to the hub is infinite \cite{prep}.

In general, when the unit voltage is applied to the resulting
circuit, by symmetry one can detect nodes which are at the same
voltage (e.g. the rims) and short them together without affecting
the distribution of currents in the branches, nor the effective
resistivity. In this way, by applying the standard
rules for the sum of  resistors in series and in parallel,
one can build an equivalent network for
the circuit, in terms of the total resistivity. To
better exploit the symmetry properties of the network, we consider
only even values of $g$, so that the points at a maximum distance
from the hub are topologically equivalent (an analogous relation
can be easily obtained if only odd values of $g$ are considered).
With some algebra one then obtains:
\begin{equation}
R_{eff}(g)={1\over 2^{g/2}}\sum_{k=0}^{g/2-1}{1\over
2^{k}}={1\over2^{g/2-1}}\left( 1-{1\over 2^{g/2}} \right)
\end{equation}
holding for even $g$. Hence, recalling that
$\mathbf{Z_g}_{11}=2(2^g-1)$ we derive that the main hub is
recurrent.

\section{\label{sec:Discussion} Conclusion and Discussion}
In this work we studied the random walk problem on a class of
deterministic networks exhibiting both a scale-free, $P(k)\sim
k^{-\gamma}$, and an exponential, $P(k) \sim (2/3)^k$, degree
distribution. The latter holds for a subset of nodes called rims
which are at a distance $1$ from the main node. Adopting a
generating function formalism we calculate the exact average
time $\tau$ to first reach the hub, where the
mean is taken over all possible walks connecting a site $i$ to the
hub and over all starting nodes $i$. The
leading behaviour for $\tau$ turns out to be $\tau \sim
V^{1-1/\gamma}$, where $V$ is the total number of sites making up
the network. Analogous power-law behaviours were found previously
for exactly decimable fractals \cite{kozak,agliari} as well as for
other deterministic scale-free networks \cite{zhang} and
Apollonian networks \cite{zhang2}, however the structure
considered here results remarkably effective, being
$1-1/\gamma \approx 0.37$. The reason lays in the large number of cycles and in the
multi-fractal nature of the the graph under study which determines
a very short average distance from the main hub.

In the second part of the work we focused on the probability to first return to the main hub and we obtained a recursive formula for its generating
function which allows to get some insight into the local topological
properties of the hub. In particular, the recurrence of the
hub is proved by mapping the graph into an electrical
network whose effective resistivity provides the escape
probability from the hub. Hence, in the thermodynamic limit, a splitting between average and local properties occurs: transience in the average is expected due to the diverging dimensionality, while the hub is locally recurrent.

Finally, we compared the return probability to the main hub with
the return probability to one of the outer nodes evidencing
different asymptotic behaviors. This further highlights the strong inhomogeneity of the graph considered.

\section*{Acknowledgments}
The authors would like to thank Olivier B\'{e}nichou, Raphael Voituriez and Alessandro Vezzani for useful and interesting discussions during the late stage of the work.

\end{document}